\documentstyle[12pt,aasms4,psfig]{article}
\newcommand{\einstein}{{\it Einstein}\ }

\newcommand{\lb}{$L_{\rm{B}}$}

\newcommand{\lxa}{$L_X$($<$4$r_e$)}
\newcommand{\lxb}{$L_X$(4--12$r_e$)}
\newcommand{\lxc}{$L_X$(12--30$r_e$)}

\newcommand{\re}{$r_{e}$}

\begin{document}
\title{Origin of the scatter in the X-ray luminosity of early-type galaxies
observed with ROSAT}
\author{{\sc Kyoko} {\sc Matsushita}\altaffilmark{1}}
\altaffiltext{1}{Max-Planck-Institut f\"{u}r extraterrestrische physik,
D-85740, Garching, Germany; matusita@xray.mpe.mpg.de}

\begin{abstract}
Statistical properties of X-ray luminosity and temperature are studied
for 52 early-type galaxies based on the ROSAT PSPC data.  All of the
X-ray luminous galaxies show largely extended emission with a radius
of a few times of 10$r_e$, while X-ray faint galaxies do not show such a
 component.  This leads to a division of early-type galaxies into two
categories: X-ray extended and X-ray compact galaxies.  Except for a
few galaxies in dense cluster environments, the luminosity and
temperature of X-ray compact galaxies are well explained by a
kinematical heating of the gas supplied by stellar mass loss.  In
contrast, X-ray extended galaxies indicate large scatter in the X-ray
luminosity. We discuss that X-ray extended galaxies are the central
objects of large potential structures, and the presence and absence of
this potential is the main origin of the large scatter in the X-ray
luminosity.

\end{abstract}

\section{Introduction}

Giant early-type galaxies are luminous X-ray emitters (e.g.\ Forman et
al.\ 1985, Trinchieri et al.\ 1986).  Studies from the \einstein
observatory showed the origin of X-rays to be hot interstellar medium
(ISM) with a temperature of $\sim 1$ keV and discrete sources such as
 low-mass X-ray binaries (LMXBs).  The hot ISM is considered
to be gravitationally confined in a galaxy, X-ray observations provide
an efficient method to look into the shape of the gravitational
potential which traces the dark matter distribution.

An additional very soft component (VSC) is often seen in the X-ray
spectra of very faint early-type galaxies, characterized by a
temperature of 0.1--0.3 keV (Kim et al.\ 1992; Fabbiano et al.\ 1994;
Pellegrini et al.\ 1994; Irwin and Sarazin 1998a).  The nature of VSC is not clear yet.

X-ray luminosities ($L_{\rm X}$) of early-type galaxies vary by two
orders of magnitude for the same optical B-band luminosity ($L_{\rm
B}$) (Trinchieri et al. 1985, Canizares et al.\ 1987, Fabbiano et al.\ 1992).  Beuing et al.\
(1999) confirmed the same $L_{\rm X}$ scatter based on the ROSAT
all-sky survey.  On the other hand, the  existence of the fundamental
plane indicates that elliptical galaxies are dynamically uniform
systems (e.g Djorgovski,  \& Davis 1987; Bender et al. 1993).
It is,
therefore, quite difficult to understand why X-ray luminosity of ISM
scatters so largely.

Various explanations for the origin of this $L_{\rm X}$ scatter have
been proposed.  Eskridge et al.\ (1995) discovered a strong
correlation between $L_X/L_B$ and potential depth, or stellar velocity
dispersion. 
White and Sarazin (1991) found that galaxies in a high
local galaxy density tend to be X-ray faint.
They conclude that the interaction with other galaxies or
surrounding ICM (intra-cluster medium) strips the ISM from the
galaxies.

Extended X-ray emission has been detected from several X-ray luminous
early-type galaxies (Trinchieri et al. 1994; Trinchieri et al.\ 1997).
Mathews et al.\ (1998) showed that X-ray luminous galaxies maintain
larger emission region than X-ray faint galaxies.  With ASCA, Matsushita et
al.\ (1998) discovered a greatly extended X-ray emission up to 300 kpc
around NGC 4636. The gravitational mass profile of NGC 4636 
indicate that the galaxy lies in the bottom of a
group-scale potential structure filled with a tenuous hot plasma.
Such extended X-ray emission has been detected in other X-ray luminous
galaxies with ASCA (Matsushita 1997).
We suggest that the presence or absence of such a greatly extended
X-ray emission is the main origin for the scatter of the X-ray
luminosity (Matsushita et al.\ 1998).

To address this issue, we need information for a larger sample of
objects.  Correct estimation for the X-ray luminosity of hot ISM has
to be made by removing the contribution from discrete sources. After
this, we can investigate the systematic difference between X-ray
luminous and X-ray faint galaxies regarding the shape of the ISM
distribution.  This paper reports statistical X-ray properties of
early-type galaxies, such as luminosity, spatial distributions of
brightness and temperature, based on the ROSAT data, which offer a
much larger sample than those of ASCA.

Galaxy distances are derived assuming the Hubble constant of $H_0 =
75$ km s$^{-1}$ Mpc$^{-1}$, using Faber et al. (1989) group distance.
We adopt the `meteoritic' solar iron abundance,
Fe/H $=3.24\times 10^{-5}$ in number, given by Anders \&
Grevesse (1989).

\section{Targets and Observations}

The early-type galaxies selected from the ROSAT PSPC archival data
have total B-band magnitude ($m_B$) larger than 11.71 and B-band
luminosity ($L_B$) larger than $10^9L_{\odot}$. The
sample consists of 52 galaxies as listed in table 1.  
It includes $\sim $ 80\% of elliptical and $\sim$ 20\%
of S0 galaxies which satisfy the selection criterion, in
the Nearby Galaxy Catalog by Tully (1998). Nineteen galaxies are
located in clusters, including 3 cD galaxies in the Virgo, Fornax and
Centaurus clusters: which are NGC 1399, NGC 4486, and NGC 4696.  
The remaining 33 galaxies are either in the field or in small groups.  The 
galaxies in the above 3 clusters are called cluster galaxies, and others are
called field galaxies, hereafter.

\section{Analysis and Results}

\subsection{Integration region and background}

The PSPC data were accumulated in 3 annular regions, 0--4\re, 4--12\re, 
and 12--30\re, centered on each galaxy. Here \re ~ is the
effective radius of each galaxy (from the \rm{RC3}
Catalog). Background is taken in a nearby outer region for each data
set. An annular region with 12--14\re~ and 30--40\re~
is chosen for the data in 0--12\re~ and 12--30\re, respectively.  This
procedure gives an automatic correction for an extended emission
component such
as an ICM contribution.  We also exclude regions within $1'$ from
discrete point sources when their signal-to-noise ratios are greater
than 4.

The PSPC spectra within 4\re~ are fitted with a standard two
hard component model: the 
Raymond \& Smith (1977) thin thermal emission for the ISM, and thermal
bremsstrahlung with a fixed temperature at 10 keV which models the
contribution from LMXBs (Matsushita et al.\ 1994). 
The two components
are allowed to have free normalization, but are subjected to a common
absorption by a free column density, $N_{\rm{H}}$.
For faint galaxies,
with the ISM flux in 4--12\re~ or in 12--30\re~ less than
$3\times10^{-13}\rm{erg~cm^{-2}s^{-1}}$, we fixed the ISM temperature
at the best-fit value obtained within 4\re. Abundances of heavy
elements are organized into two groups: $\alpha$-elements (O, Ne, Mg,
Si and S) and Fe group (Fe and Ni). Since PSPC data are not very
sensitive to metal abundance, the $\alpha$-element abundance in the
thermal component is assumed to be 1 solar following the extensive
study by Matsushita et al.\ (2000).  For every spectrum,
when  the ISM component flux  is lower than $1.5\times 10^{-12}\rm{erg~
cm^{-2}s^{-1}}$, Fe abundance is fixed to 0.5 solar.

Figure 1 shows the PSPC spectra for several representative galaxies
with different X-ray luminosities, fitted with the two component
model.  The soft ISM emission and the hard bremsstrahlung component
are distinguishable, and a single temperature model obviously gives
wrong values of ISM luminosity and temperature. The discrepancy would
be larger for X-ray faint galaxies, whose emission is more dominated
by LMXBs.  Thus, the two component model is necessary in separating
the pure ISM component.

We summarize the spectral results in table 2.  Most of the galaxies
indicate an absorption which is consistent with the Galactic level,
within $10^{20}\rm{cm^{-2}}$. 


\subsection{The ISM luminosity}
\subsubsection{X-ray extended and X-ray compact galaxies}
Figure 2 shows the  derived the ISM luminosities in the energy band 0.2-2.0 keV for
3 annular regions: within 4 \re ($L_X$($<4$\re)), 4--12 \re
($L_X$(4--12\re)), and 12--30 \re ($L_X$(12--30\re)).  
The derived luminosity values are absorption-corrected.
Significant X-ray emission from ISM is detected for 43
galaxies, and the remaining 9 galaxies give upper limits.  The
measured range of $L_X$($<$4\re) is from $10^{39}\rm{erg~ s^{-1}}$ to
$10^{43}\rm{erg~ s^{-1}}$. 
The ISM
emission in the outer 4--12\re~ and 12--30\re~ regions 
is detected for only 15 and 9 galaxies, respectively.

For X-ray
luminous galaxies, the ISM luminosity values are mostly consistent
with the published results with some difference due to the selection
of accumulating region (Trinchieri et al. 1985, Canizares et al. 1987,
Fabbiano et al. 1992, Irwin \& Sarazin 1998b, Beuing et al. 1999).
Our results for X-ray fainter galaxies are siginicantly
smaller than the earlier ones, since previous values mostly include
contribution from the hard spectral component.
Also, the central
luminosities, $L_X$($<$4\re), are consistent with the ASCA results
(Matsushita et al.\ 2000), when the difference in the energy range is
corrected. 


X-ray luminous (\lxa$>10^{41} \rm{erg~ s^{-1}}$) galaxies show
comparable levels of $L_X$(4--12\re) and $L_X$(12--30\re), and they
correlate well with the central $L_X$($<4$\re).  This correlation is
less clear for X-ray fainter galaxies mainly because of the large
uncertainty.

The spatial extent of the X-ray emission is parametrized using a ratio
of the annular luminosities. The ratios $L_X$(4--12\re)/$L_X$($<$4\re)
and $L_X$(12--30\re)/$L_X$($<$4\re) are denoted as S4--12 and S12--30,
respectively, and larger values indicate more extended emission.
Figure 3 shows S4--12 and S12--30 values against $L_X$($<$4\re). 
All of the X-ray luminous (\lxa$>10^{41} \rm{erg~ s^{-1}}$) galaxies  have
S4--12 and S12--30 to be 0.3--1.3 and 0.5--2, respectively.  If the
ISM brightness profile follows the optical $r^{1/4}$ law, these values
should be 0.13 and 0.03, respectively.  Thus, X-ray luminous galaxies
clearly show much more extended emissions than the optical profile.

In contrast, all of the S12--30 of galaxies with 
\lxa $<10^{40.0-40.8} \rm{erg~ s^{-1}}$
are less than $\sim0.5$, although 3 galaxies,
NGC 1395, NGC 3607, and NGC 4382,
have comparable level of S4--12 as  in the X-ray luminous galaxies.
Therefore, the X-ray fainter galaxies tend to have
more compact emission.


Hereafter, we denote these X-ray extended galaxies, whose S12--30 are
larger than 0.4, as ``X-ray extended'' galaxies and others as ``X-ray
compact'' galaxies.  The objects classified as X-ray extended galaxies
are NGC 1399 (cD of Fornax), NGC 1407(Group center), NGC 4406(Virgo),
NGC 4472 (Virgo), NGC 4486 (cD of Virgo), NGC 4636 (Virgo), NGC 4696
(cD of Cen), NGC 5044 (Group center), and NGC 5846 (Group center).
All of these galaxies are elliptical galaxies

\subsubsection{The scatter of X-ray luminosities against optical
properties}

The relation of $L_X$($<$4\re) to optical B-band luminosity (\lb) is
shown in Figure 4. The scatter of the ISM luminosities within the
given optical radii is two orders of magnitude for the same optical
luminosity. It is clear that the X-ray extended galaxies show larger
$L_X/L_B$ values than the  X-ray compact ones, and the division into these
two categories significantly reduces the $L_X$ scatter particularly
for the X-ray compact galaxies.

For quantitative evaluation, we fit the distributions in Figure 4.
We employ the same method in fitting the
X-ray luminosity distribution as the one for the Einstein data
analysis (Fabbiano \& Trinchieri 1985, Trinchieri \& Fabbiano 1985,
Canizares et al.\ 1987).  The distribution function is assumed to be
Gaussian with a fixed standard deviation $s$ around a straight line;
 $ \log L_X = a  (\log L_B -11.0) + b$, where $a$, $b$ and $s$ are free
parameters.  We use the maximum-likelihood method developed by Anvi
(1980), 
in which both detections and upper limits are assumed
to come from the same parent population.

The fitting results are summarized in Table 3.  The whole sample
gives the standard deviation $s$ to be 0.7, which is larger than 0.48
derived from the Einstein sample (Canizares et al. 1997).  
This is because we have separated
the pure ISM component from the rather constant discrete-source
contribution.  The best-fit slope is 1.9,
which is similar to 1.7 and 2.2 derived from the Einstein (Canizares
et al. 1987) and ROSAT (Beuing et al. 1999) data, respectively.

The deviation $s$ is reduced to 0.40 and 0.45  for the 
X-ray extended and the X-ray compact galaxies, respectively.
This indicates that the large scatter of the X-ray
luminosity is mainly caused by existence  of the largely
extended X-ray emission. 
For the X-ray compact galaxies, the slope parameter also drops to 1.5.

Figure 4 also shows correlation between \lxa~ against $L_B\sigma^2$, with
$\sigma^2$ denoting a central stellar velocity dispersion in each
galaxy.  We adopt $\sigma$ mainly from Prugniel and Simien (1996) and
also from Faber et al.\ (1985) and Whitemore et al.\ (1985). For a
given $L_B\sigma^2$, the X-ray extended galaxies show systematically
larger $L_X$ than the X-ray compact galaxies. 
The distribution against
$L_B\sigma^2$ is fitted in the same way, and the deviation $s$ for the
whole sample is 0.67. This is reduced to 0.37 by the exclusion of
the X-ray extended galaxies. Some of the X-ray compact galaxies in
clusters such as NGC 1404 and NGC 4365 show significant deviation from
the linear relation. For the X-ray compact galaxies in clusters,
the deviation $s$ is 0.52.
Thus, if we take only the X-ray compact field
galaxies, it becomes 0.28.
The best-fit slope for the
$L_X - L_B\sigma^2$ relation for the X-ray compact field galaxies is 1.3.
No difference is seen between
the X-ray compact elliptical and S0 galaxies; the average value of $\log (L_X/{L_B
\sigma^2})$ for field elliptical and S0 galaxies are 24.85 (24.74-24.94)
and 24.90 (24.70-25.05), respectively.

If stellar motion is the main heat source for the hot ISM, its X-ray
luminosity should be approximated by the input rate of the kinetic
energy of the gas from stellar mass loss. This is proportional to
$L_B\sigma^2$, since mass loss rate is thought to be proportional to
$L_B$ (e.g.\ Ciotti et al.\ 1991).  Figure 4 also shows expected
energy input from the stellar mass loss, assuming a mass loss rate
of $1.5\times10^{-11} (L_B/L_\odot) M_\odot$ yr$^{-1}$ (Ciotti et
al.\ 1991).  The expected value looks like an upper envelope to the
X-ray compact galaxies.
The ISM luminosity of those with large
$L_B\sigma^2$ agrees well with this estimation, and fainter galaxies
show slightly smaller values than the estimated level.  Only 1
X-ray compact galaxy, NGC 1404 shows much higher luminosity than the
relation. This galaxy is located near the center of the Fornax
cluster, and therefore, interaction with the surrounding ICM may raise
the ISM luminosity.  These results support the view that the ISM in the X-ray
compact galaxies is mainly heated by stellar motion. The input kinetic
energy is probably balanced with the radiation loss in the X-ray
emission, and a steady state is maintained.


\subsection{ISM temperature}
\subsubsection{Inner region}

The obtained temperatures are mostly consistent with the
previous results by various authors (Fabbiano et al. 1994, Pellegrini 
and Fabbiano, 1994, Trinchieri et al. 1997, Davis \& White 1996,
 Irwin \& Sarazin 1998b).
Since Davis \& White (1996) uses a 1 component spectral model,
our values of the X-ray faintest galaxies are systematically lower,
  and consistent to those of the VSC
(Kim et al.\ 1992; Fabbiano et al.\ 1994;
Pellegrini et al.\ 1994; Irwin and Sarazin 1998a).

Figure 5 shows correlation between ISM temperature measured within 4\re~
 and central stellar velocity dispersion, $\sigma^2$. In the
correlation for the X-ray compact objects, the cluster objects show somewhat
larger scatter than the field ones. The best-fit correlation for the
the X-ray compact field elliptical galaxies is given by $kT \propto \sigma^{1.8
(+0.4, -0.4)}$ and shown with the solid line. Thus, the ISM
temperature is  proportional to the stellar velocity dispersion.
The parameter $\beta_{\rm{spec}}$ denotes the ratio of stellar
velocity dispersion to gas temperature, i.e.\ $\beta_{\rm spec} =
\frac{\mu m_p \sigma^2}{kT}$\@ with $\mu$ indicating a mean molecular
weight in terms of the proton mass $m_p$. The plot shows that
$\beta_{\rm spec}$ is about unity for these galaxies.  Therefore, the
ISM temperature in the X-ray compact galaxies is consistent with the
kinetic energy of stellar motion.

The X-ray extended galaxies systematically exhibit a higher ISM
temperature than the X-ray compact galaxies, indicating
$\beta_{\rm{spec}}$ to be typically $\sim 0.5$.  The correlation
between $ \sigma^2 $ and $kT_s$ derived by Matsumoto et al.\ (1997),
by Davis and White (1996) show a flatter
slope, probably due to inclusion of both categories of galaxies in
their samples.


\subsubsection{ISM temperature profile}
In order to look into a temperature structure, additional spectral fits
 are carried out for the data within 1.5\re~ and 1.5--4\re. An
adjacent outer region is chosen for the background: 1.5--3\re~ and
4--6\re, respectively.

The problem in the spectral analysis is that the PSF for the PSPC is
energy dependent (Boese 1999). 
For example, when the energy is 0.2 keV, only $\sim 60\%$ of the photons from
a point source at the on-axis is detected within $0.5'$. This fraction
is more than 80\% for an energy range 0.5--1.7 keV. The accumulation
radius of such a small angular size would result in a wrong spectral
feature in the sense that the inner region shows a harder spectrum.
When the radius is raised to $1'$, the escaping photons to the outer
annulus drop to 3\% even at 0.2 keV.  We therefore exclude objects
which have $1.5 r_e < 1'$ for on-axis observations.  The PSF also
becomes broader in the outer regions of the PSPC field.  When the
source position is offset, we exclude it when the fraction of escaping
photons from $1.5 r_e$ is larger than 5\% at 0.2 keV.  As a result, 14 galaxies
are excluded. The fitting results are summarized in table 2.

%

Figure 6 compares the ISM temperatures in the central ($r<1.5r_e$) and
outer ($r>1.5r_e$) regions.  The X-ray extended galaxies all show  
positive temperature gradient with average values of
$kT(1.5-4r_e)/kT(<1.5r_e)$, $kT(4-12r_e)/kT(<1.5r_e)$, and
$kT(12-30r_e)/kT(<1.5r_e)$ to be 1.28(1.27--1.29), 1.45(1.42--1.47)
and 1.40(1.37--1.43), respectively.  The X-ray compact galaxies show
an average $kT (1.5-4r_e)/kT (< 1.5 r_e)$ to be 0.82--1.03, with
fairly large uncertainty in each data set.  Thus, even within $4r_e$,
the temperature gradient is significantly different between the X-ray
extended and the X-ray compact galaxies.

Figure 7 compares the ISM temperature against $\sigma^2$ at the galaxy
center.  As shown in \S3.3.1, values of $\beta_{\rm{spec}}$ for the X-ray
compact galaxies are $\sim$ 1.0.  
In contrast, 
The X-ray extended galaxies indicate $\beta_{\rm{spec}}=0.5-1.0$ within
$1.5r_e$ and it drops to $\sim$0.3--0.5 in $1.5r_e<r<4r_e$.
Thus, at a given central $\sigma$, 
 the temperature of the X-ray extended galaxies are systematically larger
but the differences between the two categories are relatively small at
the center.

\section{Discussion}

\subsection{Origin of VSC}

The PSPC spectra for  the galaxies are well fitted with the two
component model comprizing a soft  thin thermal emission with a temperature
0.1--1.0 keV and a hard bremsstrahlung component. The reported
spectrum of VSC (e.g.\ Kim et al.\ 1992) has normalization and
temperature consistent with our soft component.  

If the VSC is due to stellar population, we expect their X-ray luminosity
is proportional to the optical luminosity.
In fact, the X-ray luminosity of the hard component show rather good
correlation with $L_{\rm{B}}$ (Matsushita et al.\ 1994, 
Matsumoto et al.\ 1997, Matsushita et al.\ 2000).
This result suggests that  the hard component is mainly emitted
from binary X-ray sources in the host galaxy, particularly
low-mass X-ray binaries (LMXBs).
However, even in the X-ray compact galaxies,
there is still considerable scatter in the soft
component luminosity against $L_{\rm{B}}$.

As shown in \S3.3, in the X-ray compact galaxies,
the $\beta_{\rm{spec}}$ of the X-ray compact galaxies are about unity.
Also  their soft component luminosity shows a good correlation with 
a optical property, $L_B\sigma^2$.
It is also established that the main source of the X-ray emission in
luminous  galaxies is the hot ISM (e.g.\ Forman et al.\ 1986,
Trinchieri et al.\ 1987, Fabbiano 1989, Arimoto et al.\ 1997,
Matsushita et al.\ 2000).  These facts support the 
interpretation that the VSC
comes from the ISM, which is confined by the gravitational
potential. The VSC probably corresponds to the case that the soft
component temperature is very low, $< 0.3$ keV.

\subsection{ISM Mass}

The ISM mass within 4$r_e$ is calculated by assuming the ISM
brightness profile to be the same as the optical one.  
Within the optical radii, the measured X-ray 
 profile, in fact, well agrees with the optical one for both X-ray
extended galaxies, such as NGC 4636, NGC 4472, and NGC 1399 etc.\
(Forman et al.\ 1985, Trinchieri et al.\ 1994), and the X-ray compact galaxies
including NGC 720 etc.\ (Buote et al.\ 1999). 
The obtained ISM masses within 4$r_e$ are plotted in figure 8.  The X-ray
compact galaxies indicate the mass to be $10^{8\sim9} M_\odot$, which
is about 0.1\% of the stellar mass.  The X-ray extended galaxies show
$10^{9\sim10} M_\odot$, about 1\% of the stellar mass.

\subsection{ISM luminosity of the X-ray compact galaxies; heating by stellar motion}

When some galaxies in dense cluster environments are excluded, the ISM
luminosity of the X-ray compact objects is consistent with the energy
input from stellar mass loss (figure 4).  The temperature of the ISM
is consistent with the heating due to stellar motion, namely
$\beta_{\rm{spec}}\sim 1$.  The ISM abundances derived from ASCA data
are consistent with stellar metallicity (Matsushita et al.\ 2000).  As a
result, it is reasonable to conclude that the hot ISM in the X-ray
compact galaxies is mainly supplied by the stellar mass loss.

The fact that several X-ray compact galaxies in clusters do 
not follow the general correlation
suggests that interaction between ISM and surrounding
ICM causes significant variation in the ISM luminosity.

The ISM in the X-ray compact galaxies is thought to be in a steady
state, where input energy is balanced with the radiative loss. The
total ISM mass can be accumulated in only $\sim 0.1$ Gyr by the
stellar mass loss assuming the rate in Ciotti et al. (1991).
This is the same order as the radiative cooling
time scale ($0.1-1$ Gyr) and the dynamical time scale in galaxies
($\sim 0.2$ Gyr).  Therefore, the ISM in the X-ray compact galaxies
reflects almost instantaneous balance between mass supply and outflow,
as well as the balance between heating and cooling.

\subsection{External potential structure around the X-ray
  extended galaxies}

Except a galaxy within a cluster, all of the galaxies with higher 
ISM luminosity than the stellar heating 
have largely extended X-ray emission with a radius of a few times 
of 10$r_e$.
Some other energy source is needed in addition to the
kinematical heating due to the stellar motion.  
We found that
luminosities within the optical radii and in the outer extended region
correlate well.  This suggests that the extended component, rather
than the optical galaxy, has a strong influence on the ISM properties
in the central region.

We have shown in Matsushita et al.\ (1998) that the X-ray extended
galaxy, NGC~4636, is located at the  bottom of an extended potential 
structure which is filled with a tenuous plasma.  
 The present similarity in
brightness profile and in temperature profile among the X-ray extended
galaxies indicates that all these galaxies share essentially the same
hierarchical potential structure.  Temperature within 1.5--4\re\ in
these galaxies is about twice as  high as those in the X-ray compact
galaxies, indicating that their potential is deeper than those in
typical galaxies.  We hence conclude that the X-ray extended galaxies
are regarded as the central galaxy in a larger-scale potential
structure corresponding, e.g., to galaxy groups.

There are a few galaxies with a smaller scale extended
emission.  However, their ISM luminosity  are consistent with
the other X-ray compact galaxies.
These galaxies may not be at the bottom of a potential, or 
increment of the X-ray luminosity due to such a smaller scale emission
may be lower than the stellar heating.

Based on these results, Kodama \& Matsushita (2000) compared
optical properties and spatial extent of the X-ray emission.
They found no significant difference in the stellar populations
and suggested that the formation of stars predates the epoch
when the dichotomy of the potential structure was established.

We note that, if one only looks at the central region, both the X-ray
compact and the X-ray extended galaxies show a similar ISM
temperature. Positive temperature gradients (low in the center) are
usually interpreted in terms of cooling flows.  However, the observed
constancy in the ISM temperature suggests that the central cool
temperature in the X-ray extended galaxies mainly reflects the galaxy
potential, even though some part of the cool component may be produced
by cooling flows.

\section{Conclusion}

The large scatter in the X-ray to
optical luminosity ratio has been much reduced by the introduction of
a single phenomenological parameter; whether the galaxy is ``X-ray
extended'' or ``X-ray compact''.  
Interaction between ISM and surrounding
ICM also causes significant variation in the ISM luminosity.

The X-ray compact galaxies,
excluding several galaxies in dense ICM environments, show tight
correlation with kinematical energy input from stellar mass loss.
Thus, the large scatter must have been caused by a sample mixture in
the two categories.

\acknowledgements I would like to thank Kazuo Makishima and Takaya
Ohashi for valuable discussion.  This work was supported by the Japan
Society for the Promotion of Science (JSPS) through its Postdoctoral
Fellowship for Research Abroad and Research Fellowships for Young
Scientists.

\newpage


\centerline{
\psfig{file=figure1a.ps,width=8cm,angle=-90}
\psfig{file=figure1b.ps,width=8cm,angle=-90}
}
\centerline{
\psfig{file=figure1c.ps,width=8cm,angle=-90}
\psfig{file=figure1d.ps,width=8cm,angle=-90}
}
\figcaption[figure1a.ps;figure1b.ps;figure1c.ps;figure1d.ps]{
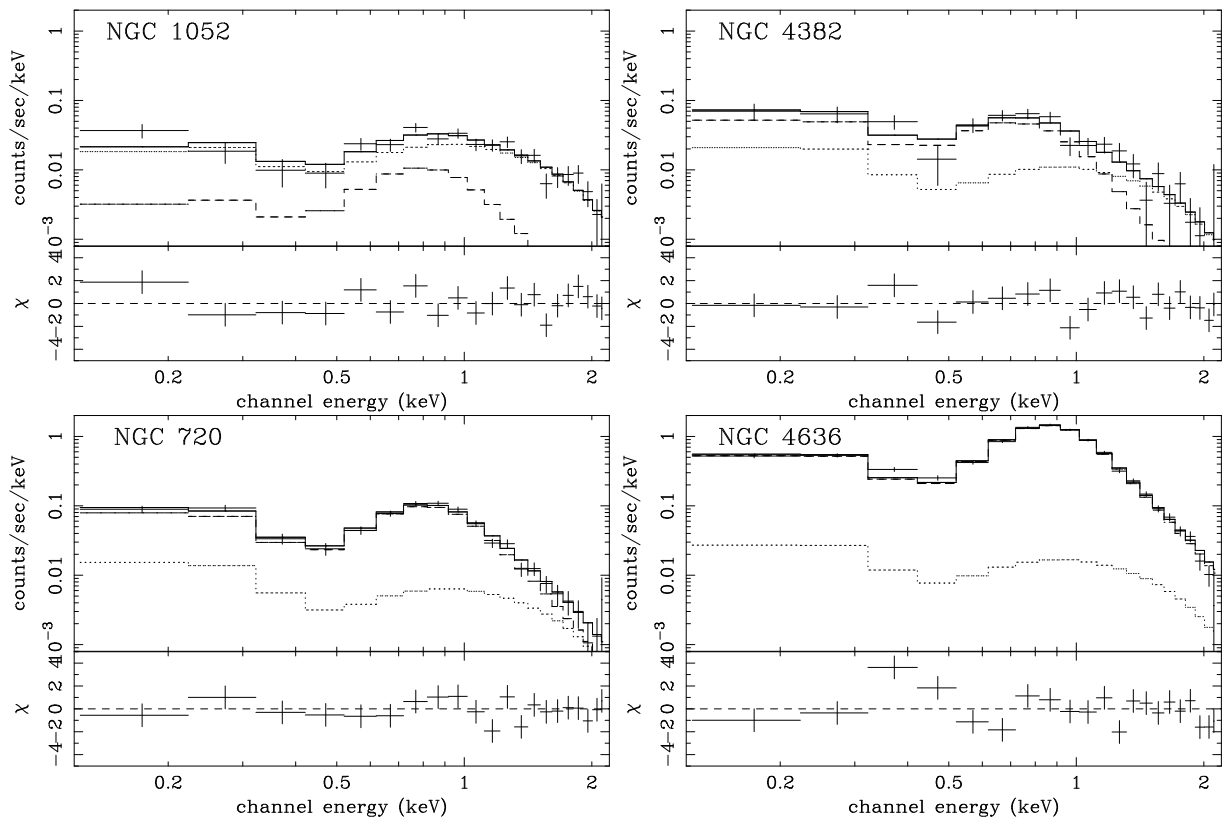
The PSPC spectra (crosses) of 4 representive galaxies 
fitted with the double-component model, consisting of a soft
R-S component (dashed line) and a hard bremsstrahlung component
(dotted line). The bottom panels show residuals of the fit.
\label{fig1}}
\clearpage

\centerline{\psfig{file=figure2.ps,width=8cm}}
\figcaption[figure2.ps]{
(upper panel) \lxb~ and (lower panel) \lxc~
of galaxies plotted against \lxa.
Marks indicate galaxy categories for X-ray extended (crosses),
 cD galaxies (large open circles),
field galaxies (open circles),  and cluster galaxies (closed circles).
\label{fig2}}
\centerline{
\psfig{file=figure3.ps,width=8cm}
}
\figcaption[figure3.ps]{
(upper panel) S4--12 (\lxb/\lxa) and (lower panel)
 S12--30 (\lxc/\lxa)  are plotted against \lxa.
Meanings of symbols are the same as figure 2.
Dotted lines corresponds to ratio of optical profile.
\label{fig3}}
\psfig{file=figure4.ps,width=16cm,angle=-90}
\figcaption[figure4.ps]{
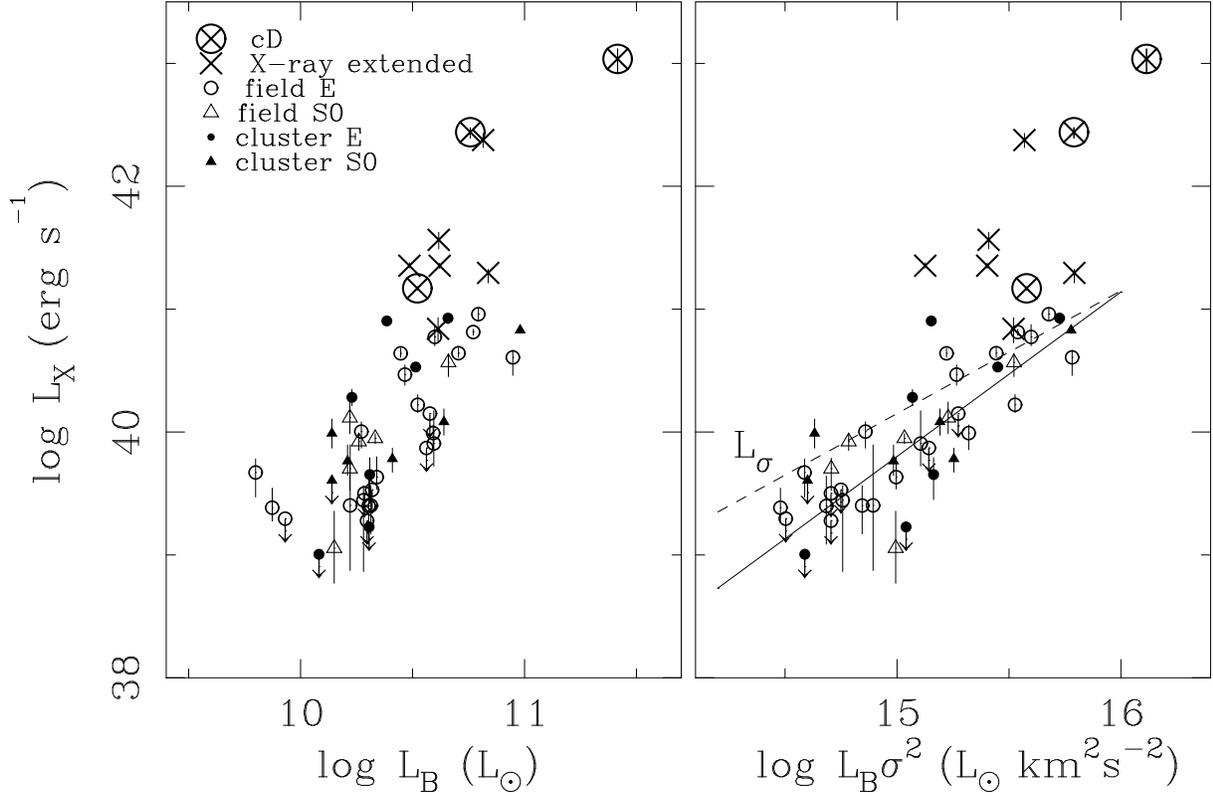
\lxa~ of galaxies plotted against 
$L_B$ (left panel) and   $L_B \sigma^2$ (right panel).  
Marks indicate
galaxy categories for X-ray extended (crosses), cD galaxies (large open circles),
field ellipticals (open circles), field S0s (open triangles), cluster ellipticals
(closed circles) and cluster S0s (closed triangles).
The dashed line  represents
 kinetic heating rate by stellar mass loss (see text in detail).
The solid line corresponds to the best fit regression line
among X-ray compact galaxies in the field.
\label{fig1}}

\psfig{file=figure5.ps,width=16cm,clip=,angle=-90}
\figcaption[figure5.ps]{
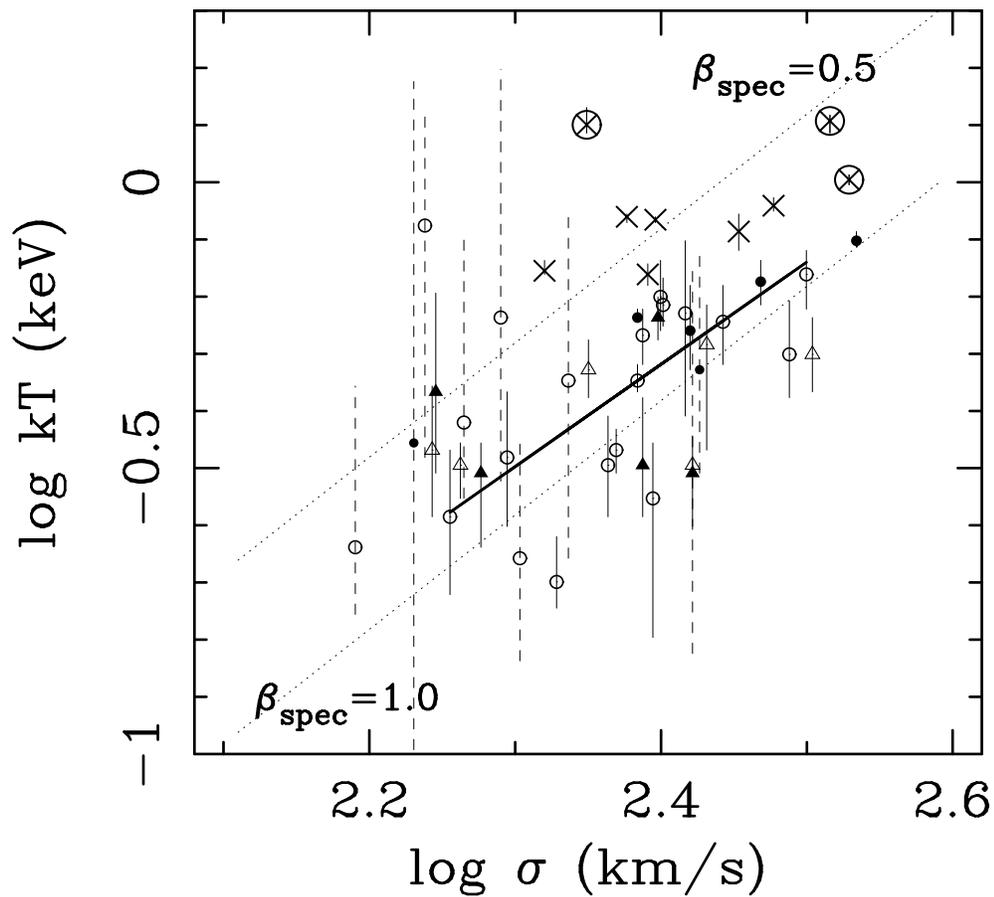
The ISM temperature within 4 $r_e$ vs. $\sigma$.
Meanings of the  symbols are the same as in figure 4.
Solid line corresponds to best fit regression line for field elliptical 
galaxies.
In order to emphasize the high quality data, the points with large
error bars are represented in dashed lines.
Dotted lines corresponds to $\beta_{\rm{spec}}=0.5$ and 1.0.
\label{fig4}}

\centerline{
\psfig{file=figure6a.ps,width=8cm,angle=-90}
\psfig{file=figure6b.ps,width=8cm,angle=-90}
}
\figcaption[figure6a.ps;figure6b.ps]{ 
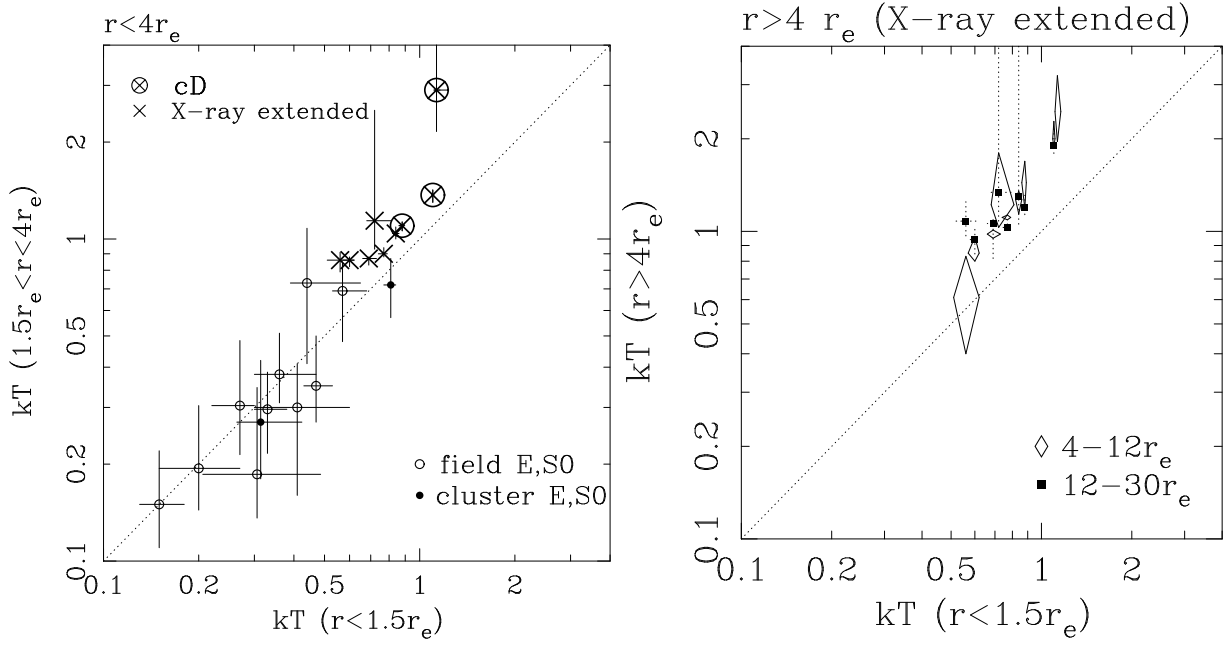
(a)  ISM Temperatures within  1.5--4 $r_e$ vs. those in 1.5 $r_e$.
Meanings of the symbols are the same as in figure 2.
Dotted line indicates the equal value for the two regions.
(b) ISM temperatures in  $r>4r_e$ vs. those in 1.5 $r_e$ 
of the X-ray extended galaxies.
Marks indicate those in  4--12 $r_e$ (diamonds),
and 12--30 $r_e$ (close squares), respectively.
 \label{fig5}}

\psfig{file=figure7.ps,width=10cm}
\figcaption[figure7.ps]{ 
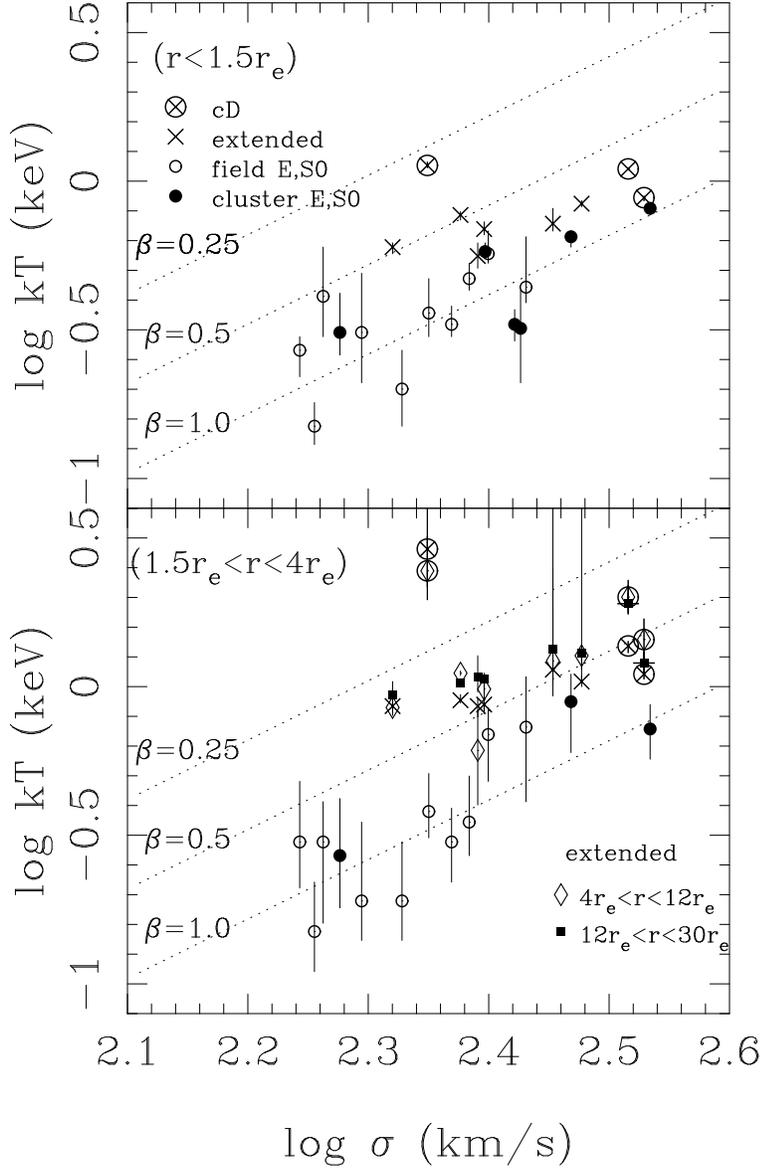
ISM temperature against $\sigma$ within 1.5 $r_e$ (upper panel), 
1.5--4 $r_e$ (lower panel).
Marks indicate
galaxy categories for X-ray extended (crosses), cD galaxies (large open circles),
field galaxies (open circles),  and cluster galaxies
(closed circles).
ISM temperature of the X-ray extended galaxies within 
4--12 $ r_e$ (diamonds) and 12--30 $r_e$ (closed squares) are
also plotted.
 \label{fig5}}

\psfig{file=figure8.ps,width=8cm}
\figcaption[figure8.ps]{ 
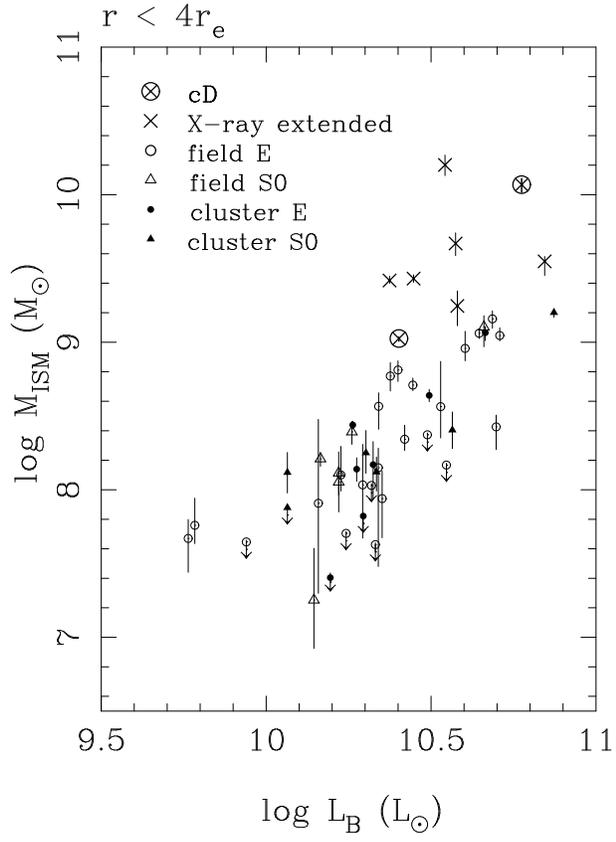
ISM mass within 4 \re~.
Meanings of the symbols are the same as in figure 4.
 \label{fig5}}
\scriptsize
\begin{table*}
\begin{center}
\caption{The galaxy sample in the ROSAT archive data}
 \begin{tabular}{rlrrrr|rlrrrr}
   galaxy & type$^a$ & $r_e^b$&exposure & offset$^c$ & note & 
   galaxy & type$^a$ & $r_e^b$&exposure & offset$^c$ & note \\
         &      & arcmin &s  & arcmin   &        &  
         &      & arcmin &s  & arcmin   &          \\\tableline
NGC 524  & -2.0 & 0.83  & 11171 & 13.7        &&NGC 4278 &-5.0  & 0.58  & 3411&0.0&\\                    
NGC 596  & -5.0 & 0.46  &  4055 & 12.6        &&NGC 4365 &-5.0 & 0.83   &14755 &0.5&Virgo \\             
NGC 720  & -5.0 & 0.60  & 22819 &  0.2        &&NGC 4374 & -5.0 & 0.85 &22020 &17.0& Virgo \\            
NGC 1052 & -5.0 & 0.56  & 13975 &  0.4        &&NGC 4382 &-2.0 & 0.91  &8495 &0.5& Virgo \\              
NGC 1316 & -2.0 & 1.35  & 25173 &0.2 & Fornax &	NGC 4406 & -5.0 &1.74  &22020&0.3& Virgo \\              
NGC 1332 & -2.0 & 0.47  & 25307  &7.1 &&	NGC 4459 & -2.0 &0.59  &7351 &14.9& Virgo \\             
NGC 1344 & -5.0 & 0.45  &  4943 &0.1&&		NGC 4472 & -5.0 &1.74  &25951 & 0.2&Virgo  \\            
NGC 1380 & -2.0 & 0.66  & 17766   & 0.8 &Fornax&NGC 4473 &-5.0 & 0.44  &7351 & 20.8 &Virgo \\            
NGC 1395 & -5.0 & 0.81  & 20464  &0.3&&	        NGC 4477 &-2.0 & 0.63  &7351 & 10.7 &Virgo  \\           
NGC 1399 & -5.0 & 0.68  & 53511 &0.2&Fornax cD&	NGC 4486 &-5.0 & 1.58  &30435 & 0.2 &Virgo cD\\          
NGC 1404 & -5.0 & 0.40  & 53511 &9.9&  Fornax &	NGC 4494 &-5.0 & 0.81  &12015 &45.2  &\\                 
NGC 1407 & -5.0 & 1.17  &22038 &0.1&& 		NGC 4526 &-2.0 & 0.74  &21354 &0.2& Virgo \\             
NGC 1549 &-5.0  & 0.78  & 15300&2.9 &&	NGC 4552 &-5.0 & 0.49  &16660 &5.2& Virgo \\              
NGC 1553 & -2.0 & 1.10  & 15300 &11.4&&		NGC 4621 &-5.0 & 0.68  &14148 &25.1&  Virgo\\            
NGC 2768 &-5.0 &  1.07  &7664 & 0.1&&		NGC 4636 &-5.0 & 1.48  &13070 &0.2& Virgo  \\             
NGC 3115 & -2.0 & 0.54  &7760 &0.4&&		NGC 4649 &-5.0 & 1.15  &14148 &1.0&Virgo \\              
NGC 3193 & -5.0 & 0.45  &4617 & 7.7&&		NGC 4696 &-4.0 & 3.53  &14580 &0.3& Cen cD \\             
NGC 3557 & -5.0 & 0.50  &19616&0.1&&		NGC 4697 &-5.0 & 1.20  &9486 &0.1&\\                     
NGC 3585 & -5.0 & 0.60  &5195& 1.6&&		NGC 5044 & -5.0 & 0.89 &27730&0.2&\\                     
NGC 3607 & -2.0 & 0.73  &32757&0.2,7.7$^d$&&	NGC 5061 &-5.0 &0.44 &6495 &0.2\\                        
NGC 3608 & -5.0 & 0.56  & 32757&12.1,6.0$^d$&&	NGC 5322 &-5.0 &0.56 &34778&0.2&                         \\
NGC 3610 & -5.0 & 0.26  & 4606&0.2&&		NGC 5846 &-5.0& 1.05 &8804&0.3\\                         
NGC 3640 & -5.0 & 0.54  &15107&0.6&&		NGC 5866 &-2.0 &0.68 &11993&0.5&                         \\
NGC 3923 &-5.0  & 0.83  &38818&0.1&&		NGC 6868 &-5.0& 0.56 &11320&27.3\\                       
NGC 4125 &-5.0 &  0.98  &10034&0.5 &&		NGC 7144 &-5.0 & 0.54 &25851&0.3&                        \\
NGC 4261 &-5.0 & 0.60   &21893&0.3&&		IC 1459 &  -5.0& 0.58 &32694&0.3\\                        \end{tabular}
\end{center} 
\tablenotetext{a}{ Morphological type code by Tully (1988)}
\tablenotetext{b}{ Effective radius  by RC3}
\tablenotetext{c}{ Offset angle  from the detector center}
\tablenotetext{d}{ Two different pointings}
\end{table*}
\tablenum{1}
\newpage

\scriptsize

\tablenum{2}
\begin{table*}
\begin{center}
 \caption{The results of spectral fitting procedure for annular regions.
Errors show 90\% confidence limits for a single parameter.}
 \begin{tabular}{r|l|rrrrrr}
   &&\tiny{Galactic} &  0--4\re & 4--12\re & 12--30\re & 0--1.5\re & 1.5--4\re\\ \hline
NGC 524 & $kT^a$ &&0.52$^{+0.09}_{-0.18}$&0.52 (fix) &0.52 (fix) &0.44$^{+0.21}_{-0.05}$&0.73$^{+0.35}_{-0.32}$\\
&$N_H^b$& 4.8&4.8 (fix)&4.8 (fix) & 4.8 (fix) & 4.8 (fix)& 4.8 (fix)\\
&${{F_X}_{\rm{ISM}}}$&&3.0$^{+0.5}_{-0.7}\times10^{-13}$&$<7.7\times 10^{-14}$ &$<1.4\times 10^{-13}$ &\\
&$\chi$$^2$/$\nu$&&19.6/18 &11.8/16 & 41.3/30  &13.8/17 & 19.0/18 \\\hline
NGC 596 &${F_X}_{\rm{ISM}}$&&$<4.6\times 10^{-14}$ &\\
&$\chi$$^2$/$\nu$&&16.0/16  \\\hline
NGC 720 & $kT$ &&0.54$^{+0.06}_{-0.06}$&0.54 (fix) &0.54 (fix)\\ &
$N_H$& 1.7&0.6$^{+0.2}_{-0.2}$&1.7 (fix) & 1.7 (fix) & \\
&${F_X}_{\rm{ISM}}$&&7.0$^{+0.7}_{-0.5}\times10^{-13}$&1.7$^{+0.9}_{-0.6}\times10^{-13}$&$<8.9\times 10^{-14}$ &\\
&$\chi$$^2$/$\nu$&&14.8/16 &24.3/33 & 21.1/20  &\\\hline 
NGC 1052 & $kT$ &&0.45$^{+0.42}_{-0.23}$&0.45 (fix) &0.45 (fix) \\
&$N_H$& 3.0&2.1$^{+1.1}_{-0.6}$&3.0 (fix) & 3.0 (fix) & \\
&${F_X}_{\rm{ISM}}$&&6.1$^{+12.9}_{-4.3}\times10^{-14}$&$<1.3\times 10^{-13}$ &$<5.7\times 10^{-14}$ &\\
&$\chi$$^2$/$\nu$&&21.4/16 &10.6/18 & 51.5/35  &\\\hline
NGC 1316 & $kT$ &&0.58$^{+0.05}_{-0.05}$&0.58 (fix) & &0.58$^{+0.04}_{-0.03}$&---$^d$\\
&Fe$^c$&&0.30$^{+0.16}_{-0.10}$ & 0.50 (fix) & & 0.50 (fix) & 0.50 (fix)\\
&$N_H$& 1.7&0.4$^{+0.2}_{-0.1}$&1.7 (fix) & & 0.8$^{+0.2}_{-0.2}$ & 1.7 (fix)  \\
&${F_X}_{\rm{ISM}}$&&1.6$^{+0.1}_{-0.1}\times10^{-12}$&$<4.6\times 10^{-13}$ &\\
&$\chi$$^2$/$\nu$&&21.1/15 &29.5/31 &  &22.8/17 & 32.6/35 \\\hline
NGC 1332 & $kT$ &&0.50$^{+0.08}_{-0.07}$&0.50 (fix) &0.50 (fix) &\\
&$N_H$& 2.1&1.2$^{+0.4}_{-0.4}$&2.1 (fix) & 2.1 (fix) & \\
&${F_X}_{\rm{ISM}}$&&3.5$^{+1.2}_{-0.9}\times10^{-13}$&$<1.0\times 10^{-13}$ &$<6.3\times 10^{-14}$ &\\
&$\chi$$^2$/$\nu$&&27.9/16 &19.7/18 & 42.1/34  &\\\hline
NGC 1344 
&${F_X}_{\rm{ISM}}$&&$<4.5\times 10^{-14}$  &\\
&$\chi$$^2$/$\nu$&&14.3/16  \\\hline
NGC 1380 & $kT$ &&0.32$^{+0.10}_{-0.06}$&0.32 (fix) & & \\
&$N_H$& 1.4&0.4$^{+0.4}_{-0.4}$&1.4 (fix) &  & \\
&${F_X}_{\rm{ISM}}$&&1.7$^{+0.6}_{-0.4}\times10^{-13}$&$<4.1\times 10^{-14}$ & \\
&$\chi$$^2$/$\nu$&&24.3/16 &20.5/18 & 
\\\hline
NGC 1395 & $kT$ &&0.63$^{+0.10}_{-0.08}$&0.51$^{+0.31}_{-0.20}$&0.63 (fix) &0.57$^{+0.11}_{-0.04}$&0.69$^{+0.16}_{-0.21}$\\
&$N_H$& 1.8&0.6$^{+0.4}_{-0.2}$&1.8 (fix) & 1.8 (fix) & 0.8$^{+0.4}_{-0.3}$& 0.0$^{+0.7}_{-0.0}$\\
&${F_X}_{\rm{ISM}}$&&5.5$^{+1.1}_{-1.0}\times10^{-13}$&3.2$^{+0.8}_{-1.2}\times10^{-13}$&$<1.1\times 10^{-13}$ &\\
&$\chi$$^2$/$\nu$&&19.7/16 &47.0/32 & 41.8/34  &9.9/17 & 14.0/17 \\\hline
NGC 1399 & $kT$ &&1.01$^{+0.02}_{-0.02}$&1.44$^{+0.25}_{-0.18}$&1.20$^{+0.11}_{-0.07}$&0.88$^{+0.01}_{-0.02}$&1.10$^{+0.03}_{-0.04}$\\
&Fe&&1.26$^{+0.42}_{-0.16}$ &1.13$^{+0.29}_{-0.15}$ & 0.93$^{+0.23}_{-0.15}$ &1.30$^{+0.51}_{-0.30}$ & 1.60$^{+1.54}_{-0.40}$\\
&$N_H$& 1.4&1.2$^{+0.2}_{-0.2}$&1.4 (fix) & 1.4 (fix) & 1.3$^{+0.2}_{-0.2}$& 0.7$^{+0.5}_{-0.4}$\\
&${F_X}_{\rm{ISM}}$&&3.5$^{+0.1}_{-0.1}\times10^{-12}$&2.0$^{+0.4}_{-0.4}\times10^{-12}$&7.5$^{+0.5}_{-0.5}\times10^{-12}$&\\
&$\chi$$^2$/$\nu$&&47.0/15 &13.4/18 & 36.3/33  &35.3/16 & 27.4/15 \\\hline
\end{tabular}
\end{center}
\end{table*}

\begin{table*}
\begin{center}
\caption{(continued)}
 \begin{tabular}{r|l|rrrrrr}
   &&\tiny{Galactic} &  0--4\re & 4--12\re & 12--30\re & 0--1.5\re & 1.5--4\re\\ \hline
NGC 1404 & $kT$ &&0.58$^{+0.02}_{-0.01}$&0.58 (fix) &&\\
&Fe&&0.59$^{+0.07}_{-0.06}$ & 0.50 (fix) &  &\\
&$N_H$& 1.4&1.0$^{+0.1}_{-0.1}$&1.4 (fix) & & \\
&${F_X}_{\rm{ISM}}$&&1.9$^{+0.1}_{-0.1}\times10^{-12}$&$<1.3\times 10^{-13}$ &\\
&$\chi$$^2$/$\nu$&&77.5/15 &41.8/30 &  &\\\hline
NGC 1407 & $kT$ &&0.82$^{+0.06}_{-0.06}$&1.22$^{+0.58}_{-0.19}$&1.34$^{+11.7}_{-0.26}$&0.72$^{+0.09}_{-0.04}$&1.14$^{+1.37}_{-0.21}$\\
&$N_H$& 5.2&4.0$^{+1.3}_{-0.9}$&5.2 (fix) & 5.2 (fix) & 5.6$^{+2.0}_{-1.0}$ & 5.2 (fix)\\
&${F_X}_{\rm{ISM}}$&&1.3$^{+0.3}_{-0.3}\times10^{-12}$&9.4$^{+2.1}_{-1.6}\times10^{-13}$&2.3$^{+0.6}_{-0.5}\times10^{-12}$&\\
&$\chi$$^2$/$\nu$&&18.8/16 &31.2/30 & 39.9/35  &13.7/17 & 12.3/18 \\\hline
NGC 1549 & $kT$ &&0.20$^{+0.04}_{-0.02}$&0.20 (fix) & 0.20 (fix) &0.20$^{+0.07}_{-0.05}$&0.19$^{+0.11}_{-0.05}$\\
&$N_H$& 1.5 &0.0$^{+1.0}_{-0.0}$&1.5 (fix) & 1.5 (fix)  & 0.1$^{+1.2}_{-0.1}$& 0.0$^{+0.7}_{-0.0}$\\
&${F_X}_{\rm{ISM}}$&&1.5$^{+0.7}_{-0.3}\times10^{-13}$&$<5.4\times 10^{-14}$ &$<4.5\times 10^{-13}$ &\\
&$\chi$$^2$/$\nu$&&26.6/16 &13.4/16 &29.9/35   &28.7/17 & 22.5/16 \\\hline
NGC 1553 & $kT$ &&0.32$^{+0.03}_{-0.04}$&0.32 (fix) &0.32 (fix) &0.41$^{+0.19}_{-0.11}$&0.30$^{+0.11}_{-0.14}$\\
&$N_H$& 1.6&0.0$^{+0.8}_{-0.0}$&1.6 (fix) & 1.6 (fix) & 0.0$^{+0.2}_{-0.0}$ & 0.0$^{+0.2}_{-0.0}$ \\
&${F_X}_{\rm{ISM}}$&&3.9$^{+0.6}_{-0.6}\times10^{-13}$&$<1.2\times 10^{-13}$ &$<5.5\times 10^{-13}$ &\\
&$\chi$$^2$/$\nu$&&26.6/16 &16.7/20 & 19.6/27  &21.9/14 & 24.4/17 \\\hline
NGC 2768 & $kT$ &&0.33$^{+0.10}_{-0.08}$&0.33 (fix) &0.33 (fix) &0.31$^{+0.18}_{-0.10}$&0.19$^{+0.16}_{-0.05}$\\
&$N_H$& 3.8&1.8$^{+2.3}_{-1.0}$&3.8 (fix) & 3.8 (fix) & 3.8 (fix)& 3.8 (fix)\\
&${F_X}_{\rm{ISM}}$&&3.0$^{+0.6}_{-0.8}\times10^{-13}$&$<1.6\times 10^{-13}$ &$<2.1\times 10^{-13}$ &\\
&$\chi$$^2$/$\nu$&&15.3/16 &30.1/33 & 39.6/33  &11.8/18 & 8.1/16 \\\hline
NGC 3115 & $kT$ &&0.32$^{+0.40}_{-0.17}$&0.32 (fix) &0.32 (fix) \\
&$N_H$& 4.5&4.5 (fix)&4.5 (fix) & 4.5 (fix) & \\
&${F_X}_{\rm{ISM}}$&&1.1$^{+1.1}_{-0.5}\times10^{-13}$&$<1.5\times 10^{-13}$ &$<6.3\times 10^{-14}$ &\\
&$\chi$$^2$/$\nu$&&23.6/16 &17.5/20 & 34.6/35  &\\\hline
NGC 3193 
&${F_X}_{\rm{ISM}}$&&$<5.7\times 10^{-14}$ &&\\
&$\chi$$^2$/$\nu$&&23.0/16 & \\\hline
NGC 3557 & $kT$ &&0.59$^{+0.20}_{-0.20}$&0.59 (fix) &0.59 (fix) &\\
&$N_H$& 8.3&8.3 (fix)&8.3 (fix) & 8.3 (fix) & \\
&${F_X}_{\rm{ISM}}$&&2.1$^{+0.3}_{-0.6}\times10^{-13}$&$<2.1\times 10^{-13}$ &$<4.5\times 10^{-14}$ &\\
&$\chi$$^2$/$\nu$&&11.0/16 &55.9/31 & 70.0/30  &\\\hline
NGC 3585 & ${F_X}_{\rm{ISM}}$&&$<3.0\times 10^{-13}$ & &\\
&$\chi$$^2$/$\nu$&&16.0/16  \\\hline
NGC 3607 & $kT$ &&0.47$^{+0.06}_{-0.05}$&0.47 (fix) &0.47 (fix) &0.36$^{+0.11}_{-0.06}$&0.38$^{+0.13}_{-0.07}$\\
&$N_H$& 1.4&0.1$^{+0.2}_{-0.1}$&1.4 (fix) & 1.4 (fix) & 0.0$^{+0.4}_{-0.0}$& 0.3$^{+1.1}_{-0.3}$ \\
&${F_X}_{\rm{ISM}}$&&3.3$^{+0.3}_{-0.3}\times10^{-13}$&2.2$^{+0.5}_{-1.3}\times10^{-13}$&$<2.0\times 10^{-13}$ &\\
&$\chi$$^2$/$\nu$&&14.9/16 &35.1/33 & 50.0/34  &18.2/17 & 24.9/17 \\\hline
\end{tabular}
\end{center}
\end{table*}
\begin{table*}
\begin{center}
\caption{(continued)}
 \begin{tabular}{r|l|rrrrrr}
   &&\tiny{Galactic} &  0--4\re & 4--12\re & 12--30\re & 0--1.5\re & 1.5--4\re\\ \hline
NGC 3608 & $kT$ &&0.22$^{+0.13}_{-0.08}$&0.22 (fix) &0.22 (fix) &\\
&$N_H$& 1.4&0.5$^{+1.1}_{-0.5}$&1.4 (fix) & 1.4 (fix) & \\
&${F_X}_{\rm{ISM}}$&&9.0$^{+4.0}_{-2.0}\times10^{-14}$&$<2.6\times 10^{-13}$ &$<3.9\times 10^{-13}$ &\\
&$\chi$$^2$/$\nu$&&13.3/16 &13.4/19 & 50.0/34  &\\\hline
NGC 3610 
&${F_X}_{\rm{ISM}}$&&$<4.4\times 10^{-14}$\\
&$\chi$$^2$/$\nu$&&9.4/16 \\\hline
NGC 3640 & $kT$ &&0.38$^{+0.42}_{-0.11}$&0.38 (fix) &0.38 (fix) &\\
&$N_H$& 6.7&6.7 (fix)&6.7 (fix) & 6.7 (fix) &\\
&${F_X}_{\rm{ISM}}$&&5.1$^{+2.3}_{-2.1}\times10^{-14}$&$<4.9\times 10^{-14}$ &$<9.0\times 10^{-14}$\\
&$\chi$$^2$/$\nu$&&13.5/17 & 16.5/17 & 13.0/16  &\\\hline
NGC 3923 & $kT$ &&0.45$^{+0.03}_{-0.02}$&0.45 (fix) &0.45 (fix) &0.47$^{+0.06}_{-0.04}$&0.35$^{+0.15}_{-0.08}$\\
&$N_H$& 6.0&6.0 (fix)&6.0 (fix) & 6.0 (fix) & 5.2$^{+2.3}_{-1.2}$& 6.0 (fix)\\
&${F_X}_{\rm{ISM}}$&&8.9$^{+0.5}_{-0.5}\times10^{-13}$&$<1.0\times 10^{-13}$ &$<5.7\times 10^{-14}$ &\\
&$\chi$$^2$/$\nu$&&13.2/16 &8.5/19 & 30.5/30  &22.1/17 & 8.4/18 \\\hline
NGC 4125 & $kT$ &&0.34$^{+0.03}_{-0.03}$&0.34 (fix) &0.34 (fix) &0.33$^{+0.05}_{-0.03}$&0.30$^{+0.09}_{-0.08}$\\
&$N_H$& 1.8&0.4$^{+0.4}_{-0.3}$&1.8 (fix) & 1.8 (fix) & 0.7$^{+0.4}_{-0.4}$&0.1$^{+0.9}_{-0.1}$\\
&${F_X}_{\rm{ISM}}$&&6.6$^{+0.7}_{-0.5}\times10^{-13}$&$<1.4\times 10^{-13}$ &$<2.6\times 10^{-13}$ &\\
&$\chi$$^2$/$\nu$&&27.2/16 &34.8/33 & 41.3/33  &40.0/35 & 29.8/35 \\\hline
NGC 4261 & $kT$ &&0.69$^{+0.07}_{-0.09}$&0.69 (fix) & \\
&$N_H$& 1.5&1.0$^{+0.2}_{-0.3}$&1.5 (fix) &  & \\
&${F_X}_{\rm{ISM}}$&&5.8$^{+1.5}_{-0.9}\times10^{-13}$&$<1.8\times 10^{-13}$ &&\\
&$\chi$$^2$/$\nu$&&12.4/16 &46.2/33 &   &\\\hline
NGC 4278 & $kT$ &&0.28$^{+0.07}_{-0.12}$&0.28 (fix) &0.28 (fix)\\
&$N_H$& 1.7&0.0$^{+1.8}_{-0.0}$&1.7 (fix) & 1.7 (fix)&\\
&${F_X}_{\rm{ISM}}$&&3.8$^{+1.1}_{-1.4}\times10^{-13}$&$<2.2\times 10^{-13}$ &$<8.0\times 10^{-13}$ \\
&$\chi$$^2$/$\nu$&&7.1/16 &10.9/16 &20.5/16  &\\\hline
NGC 4365 & $kT$ &&0.47$^{+0.29}_{-0.16}$&0.47 (fix) & 0.47 (fix) &0.32$^{+0.11}_{-0.11}$&----$^d$\\
&$N_H$& 1.6&0.5$^{+0.5}_{-0.4}$&1.6 (fix) &1.6 (fix)  & 1.4$^{+1.2}_{-0.7}$& 1.6 (fix)\\
&${F_X}_{\rm{ISM}}$&&1.6$^{+0.6}_{-0.6}\times10^{-13}$&$<2.9\times 10^{-13}$ &$<7.0\times 10^{-13}$  &\\
&$\chi$$^2$/$\nu$&&11.0/16 &16.1/18 &  &24.3/17 & 9.2/17 \\\hline
NGC 4374 & $kT$ &&0.67$^{+0.06}_{-0.06}$&0.67 (fix) & &\\
&$N_H$& 2.6&1.2$^{+0.2}_{-0.2}$&2.6 (fix) &  &\\
&${F_X}_{\rm{ISM}}$&&1.2$^{+0.1}_{-0.1}\times10^{-12}$&$<1.1\times 10^{-13}$ & &\\
&$\chi$$^2$/$\nu$&&19.1/16 &46.2/33 &  &\\\hline 
NGC 4382 & $kT$ &&0.31$^{+0.04}_{-0.08}$&0.31 (fix) &0.31 (fix) &0.31$^{+0.11}_{-0.05}$&0.27$^{+0.15}_{-0.09}$\\
&$N_H$& 2.5&0.8$^{+0.7}_{-0.5}$&2.5 (fix) & 2.5 (fix) & 1.8$^{+1.4}_{-0.9}$& 1.0$^{+1.1}_{-1.0}$\\
&${F_X}_{\rm{ISM}}$&&3.6$^{+1.0}_{-0.8}\times10^{-13}$&2.6$^{+1.7}_{-1.6}\times10^{-13}$&$<9.5\times 10^{-14}$ &\\
&$\chi$$^2$/$\nu$&&20.4/16 &28.2/32 & 50.3/37  &9.9/17 & 18.1/17 \\\hline
\end{tabular}
\end{center}
\end{table*}
\begin{table*}
\begin{center}
\caption{(continued)}
 \begin{tabular}{r|l|rrrrrr}
   &&\tiny{Galactic} &  0--4\re & 4--12\re & 12--30\re & 0--1.5\re & 1.5--4\re\\ \hline
NGC 4406 & $kT$ &&0.86$^{+0.01}_{-0.01}$&1.00$^{+0.01}_{-0.05}$&1.06$^{+0.06}_{-0.25}$&0.69$^{+0.04}_{-0.03}$&0.87$^{+0.01}_{-0.04}$\\
&Fe&&0.90$^{+0.31}_{-0.06}$& 1.0$>0.8$ & 0.30$^{+0.63}_{-0.20}$ & 0.68$^{+0.15}_{-0.14}$ & 1.20$>1.0$\\
&$N_H$& 2.6&1.7$^{+0.1}_{-0.2}$&2.6 (fix) & 2.6 (fix) & 1.4$^{+0.4}_{-0.4}$& 1.5$^{+0.4}_{-0.5}$\\
&${F_X}_{\rm{ISM}}$&&1.3$^{+0.1}_{-0.1}\times10^{-11}$&7.8$^{+0.5}_{-0.5}\times10^{-12}$&1.1$^{+0.4}_{-0.4}\times10^{-11}$\\
&$\chi$$^2$/$\nu$&&28.1/16 &26.1/16 & 36.5/33  &17.8/16 & 20.1/16 \\\hline
NGC 4459 & ${F_X}_{\rm{ISM}}$&&$<1.2\times 10^{-13}$ \\
&$\chi$$^2$/$\nu$&&13.4/16  \\\hline
NGC 4472 & $kT$ &&0.91$^{+0.03}_{-0.02}$&1.27$^{+0.08}_{-0.13}$&1.30$^{+4.10}_{-0.25}$&0.84$^{+0.02}_{-0.02}$&1.04$^{+0.05}_{-0.04}$\\
&Fe&&1.52$^{+1.20}_{-0.24}$ & 1.13$^{+1.35}_{-0.53}$ & 1.13 (fix) &1.5$^{+1.3}_{-0.4}$ & 1.8$>1.4$ \\
&$N_H$& 1.7&1.0$^{+0.1}_{-0.1}$&1.7 (fix) & 1.7 (fix) & 1.1$^{+0.2}_{-0.2}$ & 1.1$^{+0.4}_{-0.3}$ \\
&${F_X}_{\rm{ISM}}$&&7.0$^{+0.5}_{-1.2}\times10^{-12}$&3.4$^{+0.4}_{-0.4}\times10^{-12}$&3.6$^{+0.8}_{-0.6}\times10^{-12}$&\\
&$\chi$$^2$/$\nu$&&19.8/16 &32.3/36 & 31.7/35  &19.4/16 & 16.4/16 \\\hline
 NGC 4473 
&${F_X}_{\rm{ISM}}$&&$<3.6\times 10^{-14}$ \\
&$\chi$$^2$/$\nu$&&9.2/17  \\\hline
NGC 4477 & $kT$ &&0.43$^{+0.21}_{-0.12}$&0.43 (fix) &0.43 (fix)&\\
&$N_H$& 2.6&0.0$^{+0.6}_{-0.0}$&2.6 (fix) & 2.6 (fix) &\\
&${F_X}_{\rm{ISM}}$&&2.9$^{+0.9}_{-0.7}\times10^{-13}$&$<3.6\times 10^{-13}$ &$<8.5\times 10^{-13}$ \\
&$\chi$$^2$/$\nu$&&13.6/16 &21.5/33 &  25.6/23 &\\\hline
NGC 4486 & $kT$ &&1.28$^{+0.03}_{-0.06}$&2.00$^{+0.28}_{-0.12}$&1.90$^{+0.29}_{-0.13}$&1.10$^{+0.01}_{-0.01}$&1.37$^{+0.05}_{-0.07}$\\
&Fe&&$0.94^{+0.05}_{-0.08}$ & 0.94 (fix) & 0.94 (fix) & 0.97$^{+0.06}_{-0.06}$ & 1.0$^{+0.16}_{-0.13}$\\
&$N_H$& 2.6&1.5$^{+0.1}_{-0.1}$&2.6 (fix) & 2.6 (fix) & 1.3$^{+0.1}_{-0.1}$&  1.3$^{+0.1}_{-0.1}$\\
&${F_X}_{\rm{ISM}}$&&1.0$^{+0.1}_{-0.1}\times10^{-10}$&1.3$^{+0.1}_{-0.1}\times10^{-10}$&1.0$^{+0.1}_{-0.1}\times10^{-10}$&\\
&$\chi$$^2$/$\nu$&&65.9/17 &40.7/21 & 86.5/26  &59.5/16 & 20.3/16 \\\hline
NGC 4494 & $kT$ &&0.23$^{+0.21}_{-0.06}$& 0.23 (fix)\\
&$N_H$& 1.5&1.5 (fix)& 1.5(fix) \\
&${F_X}_{\rm{ISM}}$&&8.6$^{+6.4}_{-4.4}\times10^{-14}$&$<1.6\times 10^{-13}$ \\
&$\chi$$^2$/$\nu$&&13.5/17 & 15.5/17 \\\hline
NGC 4526 & $kT$ &&0.31$^{+0.05}_{-0.06}$&0.31 (fix) &0.31 (fix) &0.33$^{+0.04}_{-0.04}$&----$^d$\\
&$N_H$& 1.6&0.0$^{+1.0}_{0.0}$&1.6 (fix) & 1.6 (fix) & 1.6 (fix)& 1.6 (fix)\\
&${F_X}_{\rm{ISM}}$&&1.8$^{+0.4}_{-0.4}\times10^{-13}$&$<5.2\times 10^{-14}$ &$<1.2\times10^{-13}$ &\\
&$\chi$$^2$/$\nu$&&27.7/16 &38.4/35 & 28.4/24  &32.7/17 & 19.1/18 \\\hline
NGC 4552 & $kT$ &&0.55$^{+0.11}_{-0.08}$&0.55 (fix) & 0.55 (fix)&\\
&$N_H$& 2.6&1.7$^{+0.4}_{-0.3}$&2.6 (fix) &2.6 (fix) & \\
&${F_X}_{\rm{ISM}}$&&6.8$^{+1.1}_{-1.0}\times10^{-13}$&$<8.1\times 10^{-14}$ & $<9.0\times 10^{-14}$ &\\
&$\chi$$^2$/$\nu$&&23.3/16 &22.2/33 &27.4/27  &\\\hline
NGC 4621 
&${F_X}_{\rm{ISM}}$&&$<6.0\times 10^{-14}$ &\\
&$\chi$$^2$/$\nu$&&14.4/17 \\\hline
\end{tabular}
\end{center}
\end{table*}
\begin{table*}
\begin{center}
\caption{(continued)}
 \begin{tabular}{r|l|rrrrrr}
   &&\tiny{Galactic} &  0--4\re & 4--12\re & 12--30\re & 0--1.5\re & 1.5--4\re\\ \hline
NGC 4636 & $kT$ &&0.70$^{+0.03}_{-0.01}$&0.85$^{+0.11}_{-0.05}$&0.94$^{+0.10}_{-0.10}$&0.60$^{+0.02}_{-0.03}$&0.86$^{+0.02}_{-0.04}$\\
&Fe&&0.82$^{+0.11}_{-0.08}$ & 0.50$^{+0.15}_{-0.24}$ &0.86$^{+1.16}_{-0.45}$ &0.70$^{+0.22}_{-0.10}$ & 2.0$>1.1$ \\
&$N_H$& 1.8&1.6$^{+0.1}_{-0.1}$&1.8 (fix) & 1.8 (fix) & 1.4$^{+0.2}_{-0.1}$& 0.8$^{+0.7}_{-0.4}$ \\
&${F_X}_{\rm{ISM}}$&&8.0$^{+0.5}_{-0.3}\times10^{-12}$&3.2$^{+0.3}_{-0.2}\times10^{-12}$&8.3$^{+1.5}_{-1.3}\times10^{-12}$&\\
&$\chi$$^2$/$\nu$&&35.1/16 &37.5/33 & 65.6/50  &28.3/16 & 12.7/16 \\\hline
NGC 4649 & $kT$ &&0.79$^{+0.03}_{-0.02}$&0.79 (fix) & 0.79 (fix) &0.81$^{+0.03}_{-0.04}$&0.72$^{+0.15}_{-0.15}$\\
&Fe&&1.30$^{+3.20}_{-0.70}$ & 0.50 (fix) & 0.50 (fix) & 1.60$>1.0$ & 0.50 (fix)\\
&$N_H$& 2.0&2.4$^{+0.3}_{-0.3}$&2.0 (fix) & 2.0 (fix) & 1.6$^{+0.5}_{-0.7}$ & 1.7$^{+2.0}_{-0.8}$\\
&${F_X}_{\rm{ISM}}$&&3.0$^{+0.3}_{-0.3}\times10^{-12}$&$<4.5\times 10^{-13}$ &$<6.1\times 10^{-13}$ \\
&$\chi$$^2$/$\nu$&&25.5/16 &38.0/33 &  &10.2/15 & 13.9/17 \\\hline
NGC 4696 & $kT$ &&1.26$^{+0.09}_{-0.04}$&2.45$^{+0.76}_{-0.49}$& &1.13$^{+0.03}_{-0.02}$&2.90$^{+1.60}_{-0.74}$\\
&Fe&&0.97$^{+0.32}_{-0.14}$  & 1.0 (fix) &  & 0.84$^{+0.29}_{-0.22}$ & 1.0 (fix) \\
&$N_H$& 8.1&6.9$^{+0.5}_{-0.4}$&8.1 (fix) &  & 9.4$^{+1.2}_{-0.8}$& 8.1 (fix) \\
&${F_X}_{\rm{ISM}}$&&5.5$^{+0.9}_{-1.0}\times10^{-11}$&6.8$^{+0.4}_{-0.4}\times10^{-11}$& &\\
&$\chi$$^2$/$\nu$&&73.0/59 &17.0/17 &   &21.5/16 & 36.5/33 \\\hline
NGC 4697 & $kT$ &&0.26$^{+0.08}_{-0.07}$&0.26 (fix) &0.26 (fix) &0.15$^{+0.03}_{-0.02}$&0.15$^{+0.07}_{-0.04}$\\
&$N_H$& 2.1&0.2$^{+0.8}_{-0.2}$&2.1 (fix) & 2.1 (fix) & 2.1 (fix)& 2.1 (fix)\\
&${F_X}_{\rm{ISM}}$&&2.6$^{+2.2}_{-0.9}\times10^{-13}$&$<3.4\times 10^{-13}$ &$<1.7\times 10^{-12}$ &\\
&$\chi$$^2$/$\nu$&&30.0/16 &41.6/33 & 44.7/35  &15.7/17 & 28.6/15 \\\hline
NGC 5044 & $kT$ &&0.87$^{+0.01}_{-0.02}$&1.11$^{+0.02}_{-0.02}$&1.03$^{+0.03}_{-0.03}$&0.77$^{+0.02}_{-0.03}$&0.90$^{+0.04}_{-0.02}$\\
&Fe&&1.60$^{+0.50}_{-0.30}$ & 1.57$^{+0.15}_{-0.13}$ &0.64$^{+0.15}_{-0.12}$&1.38$^{+0.51}_{-0.40}$ & 2.44$^{+1.80}_{-0.50}$ \\
&$N_H$& 4.9&3.7$^{+0.2}_{-0.2}$&4.9 (fix) & 4.9 (fix) & 3.6$^{+0.3}_{-0.3}$& 3.5$^{+0.4}_{-0.5}$\\
&${F_X}_{\rm{ISM}}$&&1.5$^{+0.2}_{-0.2}\times10^{-11}$&8.0$^{+0.2}_{-0.2}\times10^{-12}$&9.2$^{+2.0}_{-2.2}\times10^{-12}$&\\
&$\chi$$^2$/$\nu$&&28.9/16 &19.4/17 & 49.0/30  &14.6/15 & 24.5/15 \\\hline
NGC 5061
&${F_X}_{\rm{ISM}}$&&$<1.0\times 10^{-13}$ \\
&$\chi$$^2$/$\nu$&&21.2/17 &\\\hline
NGC 5322 & $kT$ &&0.32$^{+0.07}_{-0.06}$&0.32 (fix) &0.32 (fix) &\\
&$N_H$& 1.6&0.0$^{+1.0}_{-0.0}$&1.6 (fix) & 1.6 (fix) & \\
&${F_X}_{\rm{ISM}}$&&1.2$^{+0.2}_{-0.3}\times10^{-13}$&$<1.2\times 10^{-13}$ &$<1.3\times 10^{-13}$ &\\
&$\chi$$^2$/$\nu$&&29.3/16 &17.6/19 & 35.0/30  \\\hline
NGC 5846 & $kT$ &&0.69$^{+0.03}_{-0.03}$&0.61$^{+0.22}_{-0.21}$&1.08$^{+0.19}_{-0.12}$&0.56$^{+0.06}_{-0.05}$&0.86$^{+0.05}_{-0.07}$\\
&Fe&&1.06$^{+5.54}_{-0.42}$ & 1.0$>0.40$ & 0.64$>0.14$ & 1.2$^{+1.1}_{-0.5}$ &1.6$>$0.6 \\
&$N_H$& 4.2&2.8$^{+0.6}_{-0.5}$&4.2 (fix) & 4.2 (fix) & 2.9$^{+0.9}_{-1.0}$& 2.6$^{+1.1}_{-0.7}$\\
&${F_X}_{\rm{ISM}}$&&3.8$^{+0.2}_{-0.2}\times10^{-12}$&1.3$^{+0.2}_{-0.2}\times10^{-12}$&2.8$^{+0.5}_{-0.6}\times 10^{-12}$ &\\
&$\chi$$^2$/$\nu$&&14.0/16 &46.5/33 & 32.2/25  &18.2/15 & 21.8/17 \\\hline
\end{tabular}
\end{center}
\end{table*}
\begin{table*}
\begin{center}
\caption{(continued)}
 \begin{tabular}{r|l|rrrrrr}
   &&\tiny{Galactic} &  0--4\re & 4--12\re & 12--30\re & 0--1.5\re & 1.5--4\re\\ \hline
NGC 5866 & $kT$ &&0.34$^{+0.10}_{-0.08}$&0.34 (fix) &0.34 (fix) &0.27$^{+0.03}_{-0.05}$&0.30$^{+0.18}_{-0.09}$\\
&$N_H$& 1.5&0.0$^{+0.5}_{-0.0}$&1.5 (fix) & 1.5 (fix) & 0.0$^{+2.7}_{-0.0}$& 0.0$^{+1.5}_{-0.0}$\\
&${F_X}_{\rm{ISM}}$&&1.8$^{+0.4}_{-0.6}\times10^{-13}$&$<1.3\times 10^{-13}$ &$<1.4\times 10^{-13}$ &\\
&$\chi$$^2$/$\nu$&&12.9/16 &26.8/33 & 62.7/39  &23.8/16 & 20.0/16 \\\hline
NGC 6868 & $kT$ &&0.57$^{+0.09}_{-0.09}$&0.57 (fix) & &\\
&$N_H$& 4.9&4.9 (fix)&4.9 (fix) &  &\\
&${F_X}_{\rm{ISM}}$&&5.2$^{+0.6}_{-0.6}\times10^{-13}$&$<2.7\times 10^{-13}$ &\\
&$\chi$$^2$/$\nu$&&28.6/17 &14.9/17 &  \\\hline
NGC 7144 & $kT$ &&0.84$^{+0.47}_{-0.51}$&0.84 (fix) & 0.84 (fix) \\
&$N_H$& 2.9&0.4$^{+4.2}_{-0.4}$&2.9 (fix) & 2.9 (fix) &\\
&${F_X}_{\rm{ISM}}$&&3.8$^{+1.2}_{-2.8}\times10^{-14}$&$<3.2\times 10^{-14}$ & $<5.0\times 10^{-13}$\\
&$\chi$$^2$/$\nu$&&15.5/16 &17.4/16 & 41.8/35 
\\\hline
IC 1459 & $kT$ &&0.50$^{+0.12}_{-0.08}$&0.50 (fix) &0.50 (fix) \\
&$N_H$& 1.2 & 0.6$^{+0.2}_{-0.2}$&1.2 (fix)& 1.2 (fix) \\
&${F_X}_{\rm{ISM}}$&&2.9$^{+0.6}_{-0.4}\times10^{-13}$&$<1.6\times10^{-13}$&$<8.9\times 10^{-14}$ &\\
&$\chi$$^2$/$\nu$&&10.4/16 &24.9/13 & 18.5/15  &\\\hline
\end{tabular}
\end{center}
 a: (keV)\\
 b: ($10^{20}\rm{cm^{-2}}$)\\
 c: (solar)\\
 d: not constrained
\end{table*}
\normalsize



\tablenum{3}
\begin{table*}
\caption{Fit to the distribution function. Errors show 90\% confidence limits}
 \begin{center}
 \begin{tabular}{l|cccc}
sample      & X & a  & b & $s$\\\hline
whole       & $\log(L_B)-11$   & 1.9 (1.7--2.1)  & 41.3   &0.70 (0.60--0.80)
  \\
\hline
X-ray extended& $\log(L_B)-11$ & 1.8 (1.2--2.4) & 42.5 & 0.40 (0.30--0.62)\\
compact     & $\log(L_B)-11$ & 1.5 (1.3--1.7)  & 40.8   & 0.45 (0.40--0.50) \\
  \hline
\hline
whole        &$\log(L_B\sigma^2)-16.0$   & 1.35 (1.30--1.40) & 41.4 
&0.67 (0.58--0.74)  \\\hline
X-ray extended &  $\log (L_B\sigma^2)-16.0$ & 1.6 (0.5--2.5)& 42.3 & 0.52 (0.39--0.76) \\
compact      & $\log(L_B\sigma^2)-16.0$ & 1.40 (1.20--1.60) & 41.1& 0.38 (0.32--0.46)
\\\hline
field compact & $\log(L_B\sigma^2)-16.0$ & 1.34 (1.08--1.60) &41.1 &
0.28 (0.22-0.38)\\
cluster compact & $\log (L_B\sigma^2)-16.0$ & 1.3(0.9--1.5) & 41.1 & 0.52
(0.40--0.78)\\\hline
 \end{tabular}
\end{center} 
Fits are of the form $\log L_X$ = $a X$ + $b$, with the standard diviation
 $s$.
\end{table*}

\end{document}